\documentclass[12pt]{article}

\usepackage[dvips]{graphicx}
\usepackage{epsfig}
\usepackage{amsmath,amsfonts,amssymb,amsthm}
\usepackage{verbatim}
\usepackage{psfrag}
\usepackage{bm}
\usepackage{bbm}
\usepackage{enumerate}
\usepackage{diagbox}

\usepackage{epsf,epsfig}
\usepackage{graphics}

\newcommand{\fig}{Figure}

\newcommand{\lsim}
{\;\raisebox{-.3em}{$\stackrel{\displaystyle <}{\sim}$}\;}
\newcommand{\gsim}
{\;\raisebox{-.3em}{$\stackrel{\displaystyle >}{\sim}$}\;}

\begin{document}

\thispagestyle{empty}

\begin{flushright}
{
\small
TUM-HEP-926-14
}
\end{flushright}

\vspace{0.4cm}
\begin{center}
\Large\bf\boldmath
More Viable Parameter Space for Leptogenesis
\unboldmath
\end{center}

\vspace{0.4cm}

\begin{center}
{Bj\"orn~Garbrecht}\\
\vskip0.2cm
{\it Physik Department T70, James-Franck-Stra{\ss}e,\\ 
Technische Universit\"at M\"unchen, 85748 Garching, Germany}\\
\vskip1.4cm
\end{center}

\begin{abstract}
Lepton flavour asymmetries generated at the
onset of the oscillations of sterile neutrinos with masses
above the electroweak scale can be large enough to partly
survive washout and to explain the baryon asymmetry of the Universe. 
This opens up new regions of parameter space, where Leptogenesis
is viable within the type-I seesaw framework. In particular,
we find it possible that the sterile neutrino masses are
substantially below $10^9\,{\rm GeV}$, while not being degenerate.
However, the required reheat temperature that is determined by
the begin of the oscillations lies some orders of magnitude
above the
sterile neutrino mass-scale.
\end{abstract}


\section{Introduction}

While the type-I seesaw mechanism is a very plausible way of generating
neutrino masses~\cite{Minkowski:1977sc},
it may turn out to evade experimental test.
Augmenting the Standard Model (SM)
by Majorana masses
for the active neutrinos introduces nine new parameters (three masses,
and in the PMNS matrix, three mixing angles, one Dirac and two Majorana phases),
but
there are nine extra numbers describing the masses and the couplings of the sterile neutrinos. Quite generically, the sterile neutrinos decouple from the SM
either due to their high masses or their tiny couplings, such that these
extra parameters remain practically unobservable.

Therefore, Leptogenesis~\cite{Fukugita:1986hr}
models based on the seesaw mechanism may 
likewise escape from direct tests. Nonetheless, 
the requirement of successful Leptogenesis is very
useful in order to gain more constraints on the parameter space of the seesaw
model, in particular when embedded in more general beyond the SM (BSM)
frameworks. A very well known example is the lower bound on the
masses of the sterile neutrinos, $2\times 10^9\,{\rm GeV}$~\cite{Davidson:2002qv,Buchmuller:2004nz}, which
consequently implies a lower bound of $3\times 10^9\,{\rm GeV}$
on the reheat temperature of the Universe (assuming a vanishing initial
abundance of the sterile neutrinos).

Numerous loopholes to this bound arise when considering extensions of the
SM beyond the seesaw scenario (we refer here
always to the type-I variant), see {\it e.g.} Ref.~\cite{Davidson:2008bu}.
Notable are therefore the few possibilities
that allow for smaller sterile neutrino masses and lower reheat temperatures,
but that yet rely on the minimal seesaw mechanism, thus avoiding a further
proliferation of free parameters. The most well-known
of these is resonant Leptogenesis, relying on the enhancement
of the lepton-number violating
charge-parity ($CP$) asymmetry due to the mixing of nearly
mass-degenerate sterile neutrinos~\cite{Covi:1996wh,Flanz:1996fb,Pilaftsis:1997dr,Pilaftsis:1997jf,Pilaftsis:2003gt,Garbrecht:2011aw,Garny:2011hg}.
Besides aiming for small sterile neutrino masses, it should also be of general interest to
fully chart the viable parameter space for Leptogenesis based on the
minimal seesaw mechanism.

Here, we consider Leptogenesis from
a source term that conserves lepton number but
violates the individual flavours. Its presence was first pointed
out in Ref.~\cite{Akhmedov:1998qx}, and it was subsequently
studied in a more accurate manner in
Refs.~\cite{Asaka:2005pn,Drewes:2012ma,Shuve:2014zua}. We refer to it as the
ARS scenario after the names of the authors of Ref.~\cite{Akhmedov:1998qx}.
It is based on the usual
seesaw model given by the Lagrangian
\begin{align}
\label{Lagrangian}
{\cal L}=&\frac{1}{2}\bar{N}({\rm i} \partial\!\!\!/-M) N
+\bar{\ell}{\rm i}\partial\!\!\!/\ell
+\bar{\rm R}{\rm i}\partial\!\!\!/{\rm R}
+\partial^\mu\phi^\dagger\partial_\mu \phi
\\\notag
-&\bar\ell Y^\dagger N \tilde\phi
-\bar{N} Y \ell \tilde\phi^\dagger
-\phi^\dagger \bar{\rm R} h \ell
-\phi \bar{\ell} h^\dagger  {\rm R}
\,,
\end{align}
where $\phi$ is the Higgs doublet, $\tilde\phi=(\epsilon\phi)^\dagger$  and $\epsilon$ is the ${\rm SU}(2)_{\rm L}$-invariant antisymmetric tensor.
The sterile neutrinos are given by the Majorana spinors $N_i$, for which we assume three
flavours, $i=1,2,3$, and the left- and right-handed leptons of the SM by
$\ell_a$ and ${\rm R}_a$,
where $a=e,\mu,\tau\equiv1,2,3$. We choose the usual flavour
bases for the sterile neutrinos where their mass matrix is $M={\rm diag}(M_1,M_2,M_3)$
and for the SM leptons such that
$h={\rm diag}(h_e,h_\mu,h_\tau)$. For definiteness, we
take $M_1<M_2<M_3$.

In the ARS scenario, oscillations
of the sterile neutrinos are the dominant source of flavour asymmetries (that conserve total lepton number) in the
SM leptons $\ell$ at temperatures $T\gg M_i$. The sterile neutrino masses
are typically taken to be at the ${\rm GeV}$ scale or
below~\cite{Akhmedov:1998qx,Asaka:2005pn,Drewes:2012ma}, such that
the couplings $Y$ are suppressed enough (note the seesaw
relation $m_\nu=Y^t M^{-1} Y v^2/2$, with $m_\nu$ the mass matrix
of the SM neutrinos and $v=246\,{\rm GeV}$) that
the flavoured asymmetries are only weakly washed out
prior to the electroweak phase transition (EWPT)
at the temperature $T_{\rm EW}\approx 140\,{\rm GeV}$,
where baryon-number ($B$)
violating sphaleron processes freeze out.
In Ref.~\cite{Drewes:2012ma}, it is demonstrated that in spite of the low
mass scale of the sterile neutrinos, no degeneracy is necessary in order
to account for the observed baryon asymmetry of the Universe (BAU).

We add here for clarity that in the present paper, we use the term washout
in both situations, when lepton-number violation (LNV) can be neglected for $M_i\ll T$
and when helicity flips through the Majorana masses effectively violate lepton
number once the temperature is of order $M_i$. We refer to the former case,
that yet affects the flavoured asymmetries as flavoured washout and the latter case
as LNV washout.
The case $M_i\ll T$
is relevant for the original ARS scenario, where lepton number remains effectively
conserved at times before sphalerons freeze out.
Indeed, since the sterile neutrinos are approximately chiral, one may attribute
in that situation lepton number to these particles. Transitions through
the sterile neutrinos can however lead toward chemical equilibrium between the different
flavours $\ell_a$. If chemical equilibrium is fully established,
there remain no individual asymmetries, due to complete flavoured washout. It
is essential though, that the transitions to sterile neutrinos partly take place and affect
the individual flavours in a quantitatively different manner. The
individual flavoured asymmetries are then partly hidden
from weak sphaleron transitions within the sterile neutrinos, what can result in a
net baryon asymmetry.

Here, we propose that ARS-type Leptogenesis is also viable with
sterile neutrinos above the electroweak scale, more precisely
in the mass range from $10^3\,{\rm GeV}$--$10^{11}\,{\rm GeV}$ and mass ratios
of order one.
For the subsequent considerations,
it is useful to recall the
equilibrium neutrino mass~\cite{Buchmuller:2004nz}
$m_\star=Y_\star^2 v^2/(2 M_i)$,
where $\Gamma_\star=Y_\star^2 M_i/(8\pi)$ and
$H|_{T=M_i}=\Gamma_\star$ ($H$ being the Hubble rate), such that
$m_\star\approx 1.1 \times 10^{-12}\,{\rm GeV}$.
In the present scenario, the flavour asymmetries are
produced at temperatures above the masses $M_i$ for all three $N_i$.
Given that the
observed neutrino mass-differences are larger than $m_\star$
and the large  mixing angles, the
asymmetries within each of the $\ell_{e,\mu,\tau}$
experience prior to the EWPT
strong LNV washout, defined through the condition
$\sum_i Y^\dagger_{ai}Y_{ia} M_i/(8\pi)\geq H$ ({\it i.e.} the washout rate
for $\ell_a$ exceeds the Hubble rate) at some
instance during the expansion of the Universe.
Therefore, the conversion of the flavour asymmetries
into a baryon asymmetry is less efficient than in the usual ARS scenario.
However, the mass differences of the SM neutrinos are not too far
above $m_\star$, such that there should be parametric
regions where the washout of the asymmetries is not prohibitively strong.
To demonstrate quantitatively that this is indeed a viable option,
we properly include in this work both the effects from flavoured washout and
LNV washout, that take place after the creation of the flavoured asymmetries from
oscillations but before the sphaleron freeze-out around the EWPT.
It turns out that the stronger
washout compared to the usual ARS scenario
can be made up for by larger initial asymmetries.

We mention now some similarities and differences with related papers.
In Ref.~\cite{Antusch:2009gn}, the see-saw model is constrained by certain
flavour symmetries.
As a result, $N_1$ couples to the SM particles in a way that leads to moderately strong
washout, whereas $N_{2,3}$ couple much more strongly compared to typical realisations
of the see-saw mechanism, such that large, purely flavoured asymmetries can be generated.
Moreover, the imposed flavour symmetry implies that $M_2\simeq M_3$, but it should
be noted that as it is assumed that $M_1\ll M_{2,3}$, only asymmetries from the decays
of $M_1$ are considered and the scenario does not rely on a resonance effect.
Since Leptogenesis takes place there during the decay of $M_1$ rather than
during the oscillations of sterile neutrinos, it should be clear that the scenario
from Ref.~\cite{Antusch:2009gn}
is different from what is discussed in the present paper. This is also
reflected by the  difference in the viable parameter space in terms of the $M_i$
that is found for both scenarios. (Here, the $M_i$ may be substantially smaller.)
Another variant of purely flavoured Leptogenesis with sterile neutrino masses above
the electroweak scale is discussed in Ref.~\cite{AristizabalSierra:2009bh}, where
extra particles beyond the see-saw scenario are introduced.

The outline of this paper is as follows:
In Section~\ref{sec:calculation}, we specify the way we calculate the BAU
in the proposed model and in Section~\ref{sec:ex:scenario}, we identify a viable point
in parameter space and discuss some parametric dependencies. We conclude in
Section~\ref{sec:conclusions}.

\section{Lepton Asymmetry and Washout}
\label{sec:calculation}

In this Section, we derive the equations that can be used to determine the freeze-out
asymmetry in the ARS scenario with sterile neutrinos above the electroweak scale.
While the initial flavoured asymmetries in the $\ell_a$ can be computed using the same
techniques irrespective of whether the sterile neutrino masses are above or below
the electroweak scale, the washout calculation must account for the fact that the
sterile neutrinos become non-relativistic before the sphaleron transitions freeze out,
in contrast to the original ARS proposal with sterile neutrinos below the electroweak scale.
The most important consequence of this difference is that while the sterile neutrinos
are relativistic, only flavoured asymmetries are washed out, whereas in the presence of
non-relativistic sterile neutrinos, total lepton number is effectively violated by the
Majorana masses as well.

\paragraph*{Oscillations of the sterile neutrinos and $CP$ violation.--}
We first consider the dynamics of the seesaw model when the sterile neutrinos
are relativistic, {\it i.e.} $T\gg M_i$,
and for a high reheat temperature of the SM particles.
In particular, we discuss the generation of the initial lepton asymmetry that
applies both, to the original ARS scenario and the situation with heavier
sterile neutrinos above the electroweak scale,
that is discussed for the first time in the present work.

While the Universe expands, the Higgs particle and leptons
are maintained very close to
thermal equilibrium through gauge interactions,
while the sterile neutrinos may be far from equilibrium over a wide temperature range,
because they are only interacting through the Yukawa couplings $Y$. As a plausible
initial condition, we assume that the abundance of the sterile neutrinos vanishes.
The sterile neutrinos are then produced from interactions with
leptons and Higgs bosons. For $T\gg M_{i}$, the fastest
processes involve the radiation of an extra gauge
boson ($2\leftrightarrow 2$ processes)~\cite{Anisimov:2010gy,Besak:2012qm}.
In contrast, for $T\ll M_{Ni}$,
inverse decays ($2\leftrightarrow 1$ processes) dominate.

Now, provided the matrix $YY^\dagger$ is non-diagonal, the sterile neutrinos emerge
as superpositions of their mass eigenstates and oscillate. As a result,
these coherently superimposed states
can decay into leptons $\ell$ and Higgs-bosons $\phi$ in an 
asymmetric manner~\cite{Akhmedov:1998qx,Asaka:2005pn}. This way
of creating an asymmetry is often referred to as $CP$-violating from
mixing. It is therefore
interesting to note the temperature, at which the first oscillation is
completed for a typical sterile neutrino with momentum of order of the temperature
$T$. In order to quantify this, we define
$a_{\rm R}=m_{\rm Pl}\sqrt{45/(4\pi^3 g_\star)}$, where $m_{\rm Pl}$
is the Planck mass and $g_\star$ the number of relativistic degrees of
freedom of the SM at high temperatures.
For the scale factor in the radiation-dominated Universe,
we take $a(\eta)=a_{\rm R}\eta$, where $\eta$ is the conformal time.
By this choice, $T=1/\eta$. Besides, we define $z=T_{\rm ref}/T=\eta T_{\rm ref}$.
Note that the choice
$T_{\rm ref}=M_{1}$ leads to the usual definition for the parameter
$z$ in Leptogenesis calculations~\cite{Buchmuller:2004nz}. In general,
one may find it convenient to choose $T_{\rm ref}$ to be close to the
temperature scale of interest, such that the parameter $z$ is of order one
during the relevant stages of the time-evolution. Within the final result
for the freeze-out asymmetry, the arbitrary parameter $T_{\rm ref}$ drops
out, of course.
More than one full oscillation is completed, when
\begin{align}
\label{zosc}
&\int\limits_0^\eta \frac{|M_{i}^2-M_{j}^2|}{2T} a(\eta) d\eta
=\frac{\Delta M^2 a_{\rm R} \eta^3}{6}\gsim2\pi
\\\notag
\Leftrightarrow\;\;&
z\gsim
z^{ij}_{\rm osc}=\left[12\pi T_{\rm ref}^3/(a_{\rm R}|M_{i}^2-M_{j}^2|)\right]^\frac13\,,
\end{align}
what consequently defines $T^{ij}_{\rm osc}=T_{\rm ref}/z^{ij}_{\rm osc}$
and $T_{\rm osc}=T^{13}_{\rm osc}$, as the highest of these oscillation
temperatures.
Below $T^{ij}_{\rm osc}$, the oscillations can be averaged, what
leads to a result in agreement
with the usual perturbative calculations for the decay asymmetry
of the sterile neutrinos~\cite{Covi:1996wh,Flanz:1996fb,Pilaftsis:1997dr,Pilaftsis:1997jf,Pilaftsis:2003gt,Garbrecht:2011aw}. It can be shown that when averaging the oscillations,
the time-dependence of the source term
is $\propto1/z^2$~\cite{Drewes:2012ma}. Since washout becomes important at much later
stages only, most of the asymmetry is produced around the temperature
$T_{\rm osc}$.
We recall at this point that
for the conventional lepton-number violating contributions, the
asymmetry is proportional to
$M_{i}M_{j}/|M_{i}-M_{j}|^2$, while the purely
flavoured source in the ARS scenario is enhanced by the factor
$T^2/|M_{i}-M_{j}|^2$~\cite{Drewes:2012ma}. The absence of the
chirality-flipping insertions of
Majorana mass thus explains
why the ARS-type source yields much larger asymmetries
at temperatures $T\gg M_i$ than the conventional source.
The charge density $q_{\ell a}$ in the lepton flavour $a$ that is dominantly produced
at temperatures around $T=T_{\rm osc}$  is then given by~\cite{Asaka:2005pn,Drewes:2012ma}
\begin{align}
\label{flavoured:asymmetries}
\frac{q_{\ell a}}{s}
\approx&
-\frac{1}{g_w}\sum\limits_{\overset{c}{j\not=i}}
{\rm i}
\frac
{
Y^\dagger_{ai}Y_{ic}Y^\dagger_{cj}Y_{ja}
-Y^t_{ai}Y^*_{ic}Y^t_{cj}Y^*_{ja}
}
{
{\rm sign}(M_{ii}^2-M_{jj}^2)
}
\\\notag
\times&
\left(\frac{m_{\rm Pl}^2}{|M_{ii}^2-M_{jj}^2|}\right)^{\frac 23}\times 8.4\times 10^{-5}\gamma_{\rm av}^2\,,
\end{align}
where $s$ is the entropy density.
Note that we take here a numerical factor
that is slightly smaller compared to the one of
Ref.~\cite{Drewes:2012ma},
due to a more conservative estimate of $T_{\rm osc}$, see
relations~(\ref{zosc}).

Note that Eq.~(\ref{flavoured:asymmetries}) corresponds to
a conservative estimate that may receive corrections of order one~\cite{Drewes:2012ma}.
In particular, calculating the momentum-dependent phase-space distributions
instead of solving for the number densities of the sterile neutrinos should improve the
predictions.
A complete leading order accurate  calculation appears viable and will be tackled in the 
future. One can immediately verify
that Eq.~(\ref{flavoured:asymmetries})
implies that $\sum_a q_{\ell a}=0$, because these asymmetries
are purely flavoured. As stated in the introduction, it is important that at a later stage,
the individual flavour asymmetries are affected differently by flavoured or
LNV washout, such that eventually a total lepton asymmetry within the
SM leptons arises. Note that
differently from Ref.~\cite{Drewes:2012ma}, we define $q_X$ as the
charge density within one component of the ${\rm SU}(2)$ multiplet,
hence the factor $1/g_w$ with $g_w=2$.

An important input parameter to Eq.~(\ref{flavoured:asymmetries}) is $\gamma_{\rm av}$,
the average production rate of sterile neutrinos. It is derived from the total production rate for
$N_i$ per unit volume, $[YY^\dagger]_{ii}\gamma^N$, by dividing it with the number density of $N_i$:
\begin{align}
\gamma_{\rm av}=\frac{\gamma^N}{2n_N^{\rm eq}}\,,
\end{align}
where
$n_N^{\rm eq}=\int d^3k/(2\pi)^3\,f_{\rm F}^{\rm eq}(\mathbf k)$ and $f_{\rm F}^{\rm eq}(\mathbf k)$ is the Fermi-Dirac distribution of a massless fermion.
It appears
quadratic in Eq.~(\ref{flavoured:asymmetries}), because one factor can be attributed to the available
phase space and the other factor to the $CP$-violating cut.
We take the
value $\gamma^N=2.2\times 10^{-3} T^4$, where we have used $h_t=0.6$, $g_2=0.6$
and $g_1=0.4$ for the top-quark Yukawa-coupling and
for the ${\rm SU}(2)$ and ${\rm U}(1)$ gauge couplings,
what should be suitable for an energy scale of
$10^8\,{\rm GeV}$.
Consequently, $\gamma_{\rm av}=0.012$, which is larger than the
value used in Ref.~\cite{Drewes:2012ma}. The estimate there is based on a
partial evaluation of the sterile neutrino production rate in the relativistic regime~\cite{Anisimov:2010gy},
while the complete leading-order results can be taken from the more recently published Refs.~\cite{Besak:2012qm,Garbrecht:2013bia}.
Note that $\gamma^{\rm av}$ also appears in the flavoured washout
rate in the relativistic regime, Eq.~(\ref{W:rel}) below. Larger values for
$\gamma_{\rm av}$ therefore enhance both the initial asymmetry and the washout,
such that in view of the freeze-out asymmetry, theoretical uncertainties in this rate
should be partly compensated.

One may take into account a temperature
dependence due to the running coupling constants~\cite{Anisimov:2010gy,Garbrecht:2013bia}, which
we ignore here in view of the accuracy of other approximations
used in the present calculation of the freeze-out asymmetry.
In addition, we estimate the temperature $T_{\rm res}$,
above which the largest of the widths of the sterile neutrinos exceeds the mass splitting, 
such that Eq.~(\ref{flavoured:asymmetries}) should not be applied,
as
\begin{align}
T_{\rm res}=\sqrt{(M_2^2-M_1^2)/(\gamma_{\rm av}\max_i[YY^\dagger]_{ii})}\,.
\end{align}

To summarise, the asymmetries $q_{\ell a}$ given by Eq.~(\ref{flavoured:asymmetries})
are present after the initial oscillations of the sterile neutrinos at temperatures
around $T_{\rm osc}$, but before sizable flavoured and LNV washout effects occur.
We may therefore use the asymmetries~(\ref{flavoured:asymmetries}) as initial
conditions for Eqs.~(\ref{washout:equil:eqs})
or~(\ref{BEqs:withhelicityasymmetry})
that describe the evolution of the flavoured asymmetries
through washout.
The time separation between the production of the initial asymmetries and their washout
was first noted and made use of
in a calculation in Ref.~\cite{Asaka:2005pn}. To explicitly validate
this procedure within the present context,
we define the temperature $T=T_{{\rm W}a}$ at which the rate for
flavoured washout of $\ell_a$ equals the Hubble rate,
$W^{\rm rel}_a=H$ [see Eq.~(\ref{W:rel}) below], such that
\begin{align}
T_{{\rm W}a}=m_{\rm Pl}\gamma_{\rm av}\sqrt\frac{45}{4\pi^3 g_\star}[Y^\dagger Y]_{aa}\,.
\end{align}
This should be compared with $T_{\rm osc}$, and in particular for the relevant flavour $e$,
we must demand that $T_{{\rm W}e}\ll T_{\rm osc}$, for the separation of the initial production
from washout to be valid. For our numerical example this comparison is made
in \fig~\ref{fig:TOscTRes}.
We emphasise that
expression~(\ref{flavoured:asymmetries}) applies to both,
the original ARS scenario with light sterile neutrinos as well as the
scenario proposed here with sterile neutrinos above the electroweak scale, and
it can be obtained by integrating over the initial oscillations of the
sterile neutrinos. It was first derived in Ref.~\cite{Asaka:2005pn}, based
on a density matrix description of the sterile neutrinos, following the
original work~\cite{Akhmedov:1998qx}. Ref.~\cite{Drewes:2012ma} relies on
an alternative method, using a Green-function description
of the sterile neutrinos within a non-equilibrium framework. The analytic
formula~(\ref{flavoured:asymmetries}) does not only correspond to a simpler way of 
computing the asymmetry compared to numerically solving for the oscillations of the
sterile neutrinos, it also transparently exhibits the dependence of the initial asymmetry
on the model parameters\footnote{In particular, we remark that $q_{\ell a}/s$
can be of order $10^{-10}$ ({\it i.e.} of order of the observed BAU)
even when the Yukawa interactions mediating 
between $\ell_a$ and the $N_i$ are far from equilibrium, {\it i.e.} when
$[Y^\dagger Y]_{aa} T_{\rm osc}\ll T_{\rm osc}^2/m_{\rm Pl}$
or equivalently $[Y^\dagger Y]_{aa}\ll (|M_{ii}^2-M_{jj}^2|/m_{\rm Pl}^2)^{\frac13}$.}.

\paragraph*{Washout of left-handed leptons.--}
Below the temperature $T_{\rm osc}$, the asymmetry $q_{\ell e}/s$ given
by Eq.~(\ref{flavoured:asymmetries})
is approximately conserved initially until washout processes
become important\footnote{{\it Cf.} the evolution of the
particular flavours shown in \fig~\ref{fig:SterileWO} and the comparison
between $T_{{\rm W}e}$ and $T_{\rm osc}$ in \fig~\ref{fig:TOscTRes}. The temporal separation
of the initial production of the asymmetry and the washout that mostly
takes place at later stages is also made use of in Refs.~\cite{Asaka:2005pn,Drewes:2012ma,Shuve:2014zua}.}.
A net lepton asymmetry is then present at the time of sphaleron
freeze-out during the EWPT, provided
the individual lepton flavours are affected differently by the
washout. For non-relativistic sterile neutrinos, the LNV washout rate of $\ell_a$
is usually obtained from the thermally averaged inverse decay
rate, approximating the quantum statistical
by classical Maxwell distributions~\cite{Buchmuller:2004nz}.
In our parametrisation (in particular, using the dimensionless
variable $z$ as the time parameter), this quantity is given by
\begin{align}
W^{\rm NR}_a=&\sum\limits_iY^\dagger_{ai}Y_{ia}\left(\frac{M_i}{T_{\rm ref}}\right)^\frac52
\frac{a_{\rm R}}{T_{\rm ref}}\frac{3}{2^{\frac72}\pi^\frac52}
{\rm e}^{-z\frac{M_i}{T_{\rm ref}}}z^{\frac52}
=\sum\limits_i \bar W^{\rm NR}_i Y^\dagger_{ai}Y_{ia}\,,
\end{align}
while in the relativistic regime, the rate of equilibration
of $\ell_a$ with the sterile neutrinos ({\it i.e.} the flavoured washout rate) is
\begin{align}
\label{W:rel}
W_a^{\rm rel}=\sum\limits_iY^\dagger_{ai}Y_{ia}\gamma_{\rm av}
=\bar W^{\rm rel}\sum\limits_iY^\dagger_{ai}Y_{ia}\,.
\end{align}
One should note that the latter rate
is lepton-number conserving, as it does not rely on the insertion
of Majorana mass terms, which only enter at sub-leading order in
the relativistic regime. Below, we consider a region in parameter space
where $|Y_{ie}|\ll|Y_{i\mu},Y_{i\tau}|$, such that only $q_{\ell e}$
is not nullified by washout, and any asymmetry
that is transferred from $q_{\ell e}$ into relativistic sterile neutrinos will
scatter or decay into $\ell_{\mu,\tau}$.
We therefore take for the effective washout rate of $\ell_e$
\begin{align}
\label{washout:combined}
W_e=\sum\limits_i Y^\dagger_{ei}Y_{ie}\big[&\vartheta(T-M_i)\max(\bar W_i^{\rm NR},\bar W^{\rm rel})
+
\vartheta(M_i-T)\bar W_i^{\rm NR}\big]\,.
\end{align}
This accounts for the relativistic $2\leftrightarrow 2$ 
scatterings at high temperatures (flavoured washout),
for the $1\leftrightarrow 2$ decays
and inverse decays when $M_i\gsim T$ (LNV washout) and eventually for the
freeze-out of these reactions due to Maxwell suppression when
$M_i\gg T$.
Note that those Figures of Refs.~\cite{Besak:2012qm,Garbrecht:2013bia} that
show the individual relativistic lepton-number conserving and the non-relativistic LNV
contributions to the sterile neutrino production rate are well suited to illustrate the
particular reactions we account for in the combined washout rate
$W_e$, that is given by Eq.~(\ref{washout:combined}).

\paragraph*{Complete and partial chemical equilibrium of right-handed SM leptons.--}
The Yukawa couplings $h_a$ mediate chemical equilibration
between the $\ell_a$ and ${\rm R}_a$, at a rate that
is given by $h_a^2 \gamma^{\rm fl}T$
with $\gamma^{\rm fl}=5\times 10^{-3}$~\cite{Cline:1993bd,Garbrecht:2013bia}.
Full equilibration
implies $q_{\ell a}=q_{R a}+\frac12 q_H$. (Note again our convention that the charge densities
account for one component of a multiplet only. Hence, this relation follows
from the condition for the chemical potentials $\mu_{\ell a}=\mu_{{\rm R}a}+\mu_H$, where
$q_{\ell a,{\rm R}a}=\mu_{\ell a,{\rm R}a} T^2/6$ and $q_H=\mu_H T^2/3$.) Again, we ignore the
temperature dependence of $\gamma^{\rm fl}$, which is present
because of running coupling strengths.
The redistribution of the
asymmetries is quantitatively relevant,
because it reduces the washout rates, as only the
$\ell_a$ couple to the sterile neutrinos. While chemical equilibrium between $\ell_{\mu,\tau}$
and ${\rm R}_{\mu,\tau}$ can be assumed for all temperatures below $1.3\times 10^9\,{\rm GeV}$,
we account for the possibility of only partial equilibration
of  $\ell_{e}$ and $R_{\rm e}$.
We include other spectator effects following
Ref.~\cite{Barbieri:1999ma}. In particular, we impose the following conditions:
\begin{enumerate}[\label=(i)]{}
\item
\label{cond:Yukawaeq}
Chemical equilibrium is maintained by all SM Yukawa couplings but the electron-Yukawa coupling
$h_e$. For the latter, we account for partial equilibration within the Boltzmann equations.
\item
Strong and weak sphalerons are in chemical equilibrium.
\item
\label{cond:mutauwashout}
The reactions mediating the process $\ell_{\mu,\tau}+\phi\leftrightarrow N_{1,2,3}$ are in equilibrium. (This of course must be verified for each particular choice of parameters made.) Note that in the relativistic regime, where $M_i\ll T$, these reactions involve at
leading order the
radiation of extra particles for kinematic reasons. We assign no chemical potentials
to the $N_i$, as it is appropriate for the non-relativistic regime, where the Majorana
masses $M_i$ flip the sterile neutrino helicities. In the relativistic regime, one would need in principle
a more refined description, tracking the number densities of the individual
helicity states. As the washout of the initial asymmetry occurs mostly during the non-relativistic LNV regime, neglecting the finite rate of helicity flips should amount to a small correction only.
\end{enumerate}
Regarding condition~(\ref{cond:Yukawaeq}), we note that
equilibrium of $u$- and $d$-quark Yukawa mediated interactions may not
be an appropriate assumption within the higher range, above $2\times10^{6}\,{\rm GeV}$,
for the sterile neutrino masses that we
consider. Moreover, above $10^9\,{\rm GeV}$, $s$ quarks and $\mu$ leptons
should equilibrate as well. The low-mass region, that is perhaps the phenomenologically most interesting,
should however not be affected by this simplification.
For higher masses, the error incurred by not accurately treating
the chemical equilibration of these non-leptonic interactions should be at the 10\%--20\% level~\cite{Nardi:2005hs}.

In combination, above conditions force the relations
\begin{subequations}
\begin{align}
q_{\ell e}=&-\frac{11}{26}\Delta_e-\frac{157}{390}q_{{\rm R}e}\,,\\
q_H=&-\frac{1}{13}\Delta_e+\frac{7}{195}q_{{\rm R}e}\,,
\end{align}
\end{subequations}
where $\Delta_a=B/3-2q_{\ell a}-q_{{\rm R}_a}$ is the flavoured lepton asymmetry that
is conserved by weak sphaleron processes, with $B$ being the baryon charge density.
(The factor 2 is due to our convention that $q_{\ell a}$ counts the charge within
one component of the ${\rm SU}(2)$ doublet only.) In order to compute the final freeze-out
asymmetry, we note that the total baryon-minus-lepton number density is
\begin{align}
B-L=\Delta_e+\frac{8}{15}q_{{\rm R}e}\,.
\end{align}

Besides, it may be of interest to consider the effect of the partial chemical equilibration
of the right-handed electron ${\rm R}_e$ by comparing the resulting asymmetry with the outcome
in the two limiting cases of full chemical equilibrium and a vanishing chemical potential for
${\rm R}_e$. For a vanishing chemical potential, we can set $q_{{\rm R}_e}=0$ in above
relations, whereas
in the case of full equilibration, we obtain
\begin{align}
q_{\rm\ell e}=-\frac{87}{277}\Delta_e\,,\qquad q_{H}=-\frac{24}{277} \Delta_e
\end{align}
and
\begin{align}
B-L=\frac{237}{277}\Delta_e\,.
\end{align}

\paragraph*{Final baryon asymmetry.--}
Putting together above details, equations that can
be used to determine the BAU for the present scenario
are simply given by ({\it cf.} Ref.~\cite{Cline:1993bd}):
\begin{subequations}
\label{washout:equil:eqs}
\begin{align}
\frac{dq_{\Delta e}}{d z}=&2\,W_e\left( q_{\ell e}+\frac12 q_H\right)\,,
\\
\frac{d q_{{\rm R}e}}{dz}=&-2\gamma^{\rm fl}\frac{a_{\rm R}}{T_{\rm ref}}
h_e^2\left(q_{{\rm R} e}-q_{\ell e}+\frac12 q_H\right)
\,.
\end{align}
\end{subequations}
As initial conditions at $z=0$, we take $q_{{\rm R}e}=0$
and $q_{\ell e}$ obtained from Eq.~(\ref{flavoured:asymmetries}).
The solution evaluated at $z=T_{\rm ref}/T_{\rm EW}$ yields the $e$-lepton
charge density $L_e=g_w q_{\ell e}+q_{{\rm R} e}$ at the EWPT.
Due to the strong washout in the $\mu,\tau$  flavours,
there is no asymmetry in these sectors, and
we obtain for the BAU $B\approx-(28/79) L_e$~\cite{Harvey:1990qw},
which is to be compared with the observed value
$B_{\rm obs}/s=8.6\times 10^{-11}$~\cite{Hinshaw:2012aka,Ade:2013zuv}.

\section{Example Scenario}
\label{sec:ex:scenario}

\paragraph*{Freeze-out asymmetry.--}

On the case of a specific parametric reference point,
we demonstrate now  that
the proposed scenario can explain the BAU in a phenomenologically viable way.
We use the parametrisation of the neutrino Yukawa
couplings from Ref.~\cite{Casas:2001sr} and follow the notation of
Ref.~\cite{Drewes:2012ma}.
For the mixing angles of the PMNS matrix and the mass differences of the
SM neutrinos, we take the best fit values from
Ref.~\cite{Fogli:2012ua}. We assume a normal mass hierarchy with the lightest 
neutrino mass
$m_1=2.5\,{\rm meV}$. As for the sterile neutrinos, we
take $1:2:3$ for the mass ratios $M_1:M_2:M_3$. For the remaining angles,
we choose
\begin{equation}
\label{example:parameters}
\begin{aligned}
\delta=&0.2\,,\quad\\
\alpha_1=&0\,,\\
\alpha_2=&2.6\,,
\end{aligned}
\begin{aligned}
\omega_{23}=&0.6+1.4{\rm i}\,,\\
\omega_{13}=&0.1-1.5{\rm i}\,,\\
\omega_{12}=&-1.9-1.0{\rm i}\,.
\end{aligned}
\end{equation}


\begin{figure}[t!]
\begin{center}
\epsfig{file=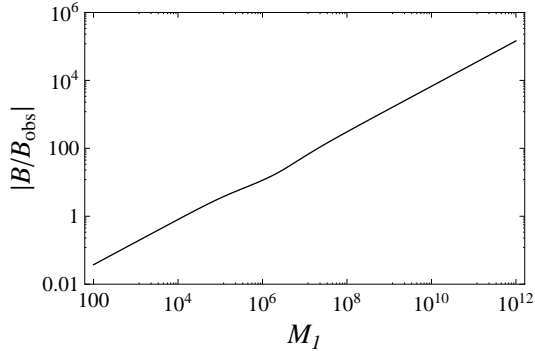,width=7cm}
\end{center}
\vskip-.4cm
\caption{
\label{fig:BAU}
Final BAU normalised to the observed value for the reference parameters.
}
\end{figure}

\begin{figure}[t!]
\begin{center}
\epsfig{file=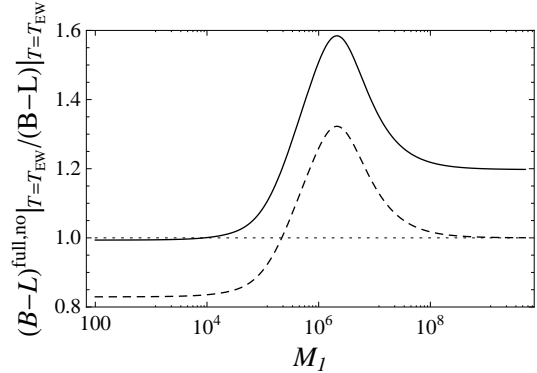,width=7cm}
\end{center}
\vskip-.4cm
\caption{
\label{fig:wiggle}
Ratio of the final BAU assuming either full equilibration (solid) or no equilibration
of ${\rm R}_e$ compared to the result assuming partial equilibration for the reference parameters. The dotted line for the ratio one should help to guide the eye and to
demonstrate that both limiting cases apply for small and large $M_1$.
}
\end{figure}

\begin{figure}[t!]
\begin{center}
\epsfig{file=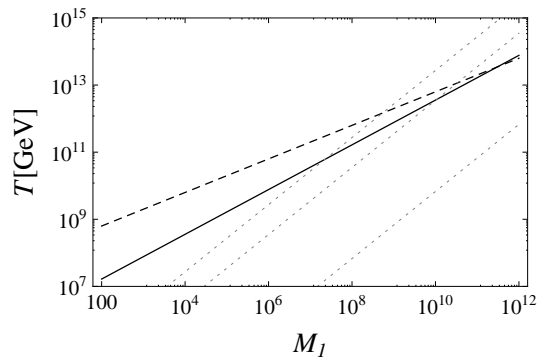,width=7cm}
\end{center}
\vskip-.4cm
\caption{
\label{fig:TOscTRes}
The temperature $T_{\rm osc}$ (solid) and
the temperature $T_{\rm res}$ (dashed)
for the reference parameters. The dotted lines represent
$T_{{\rm W}e,\mu,\tau}$ from the bottom right to top left.
}
\end{figure}

\begin{figure}[t!]
\begin{center}
\epsfig{file=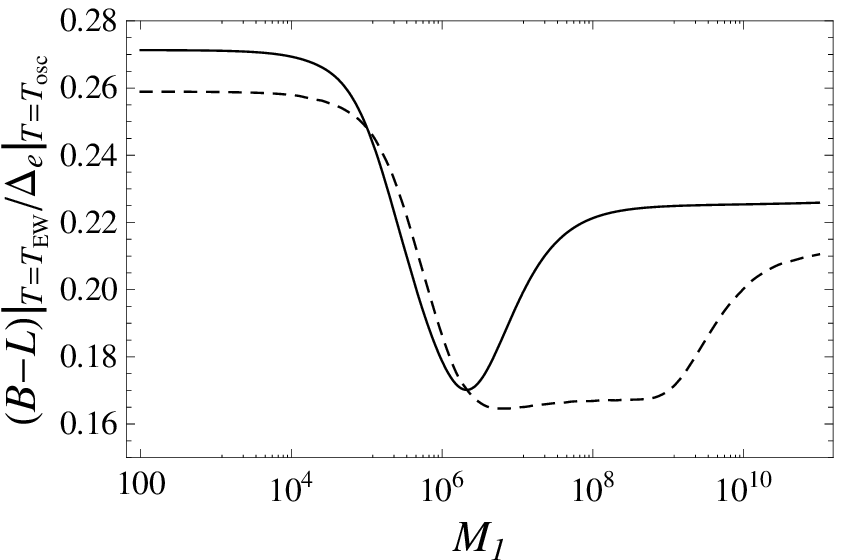,width=7cm}
\end{center}
\vskip-.4cm
\caption{
\label{fig:fraction}
Fraction of the initial asymmetry that remains after washout for the reference 
parameters. Result based on the approximations described in
Section~\ref{sec:calculation} (solid), where the
asymmetries in $q_{Ni}$ and $\Delta_{\mu,\tau}$ are neglected are compared
with the results from the equations of Appendix~\ref{appendix:helicity:asymmetries} (dashed),
where the detailed evolution of the individual flavours are taken into account.
}
\end{figure}

We vary $M_1$, keeping the mass ratios of the sterile neutrinos
fixed. In \fig~\ref{fig:BAU}, we show the ratio of the final BAU to the observed value
as a function of $M_1$. It turns out that $B/B_{\rm obs}=1$ for
$M_1\approx 4.8\times 10^3\, {\rm GeV}$ and
$T_{\rm osc}\approx 2.2\times 10^8\,{\rm GeV}$. This value for $T_{\rm osc}$
can be interpreted as the minimum reheat temperature for
the particular parametric scenario~(\ref{example:parameters}), and it happens to be somewhat below the
corresponding value $2.0\times 10^9\,{\rm GeV}$ for standard
Leptogenesis~\cite{Buchmuller:2004nz}. The increase of the
asymmetry with $M_1$ can be easily understood in terms of the initial
asymmetry~(\ref{flavoured:asymmetries}) and the seesaw relation. The deviation
from a simple power law, visible as a wiggle, is due to the
equilibration of the spectator particles ${\rm R}_e$,
described by Eqs.~(\ref{washout:equil:eqs}). This feature is highlighted
in \fig~\ref{fig:wiggle}, where we compare the asymmetries from assuming either
full or no chemical equilibration of ${\rm R}_e$ to the full result, solving
for the asymmetry within ${\rm R}_e$ through Eqs.~(\ref{washout:equil:eqs}). The
fact that the interpolation between the two limiting cases does not proceed
monotonously is because initially larger asymmetries may be effectively
hidden in the spectators before they fully equilibrate. This observation
is related to the discussion of washout of primordial asymmetries of 
Ref.~\cite{Cline:1993bd}. In the context of standard Leptogenesis, it is investigated
in Ref.~\cite{Garbrecht:2014kda}.
The mass range in which our approximations are viable is
given by the condition $T_{\rm osc}<T_{\rm res}$, and from
\fig~\ref{fig:TOscTRes}, we see that we should choose $M_1$ to be below
about $10^{11}\,{\rm GeV}$. We emphasise again that in the higher
temperature range, the spectator fields such as the $\mu$-lepton,
and $s$, $d$, $u$ quarks do not equilibrate, which we do not account of here,
as we find the low mass range phenomenologically more interesting and the
error of order $10\%$ to $20\%$ incurred~\cite{Nardi:2005hs} is comparable with other theoretical uncertainties
in the present calculation. From \fig~\ref{fig:TOscTRes}, we can also see that the approximation
that we take in separating the production of the initial asymmetry in $\ell_e$ from its washout
is met everywhere, since $T_{{\rm W},e}\ll T_{\rm osc}$.

\paragraph*{Effective washout strength.--}

Within the studied range for $M_1$, we find that
$(B-L)|_{T=T_{\rm EW}}/\Delta_e|_{T=T_{\rm osc}}$ varies between
0.2 and 0.17, {\it cf.} \fig~\ref{fig:fraction}, indicating that for the reference point, the 
$|Y_{ie}|$ takes small values such that the washout of the $\ell_e$ 
is as small as possible. We can verify this more explicitly in terms of the
usual LNV washout parameters $K_{ia}=|Y_{ia}|^2 M_i/(8\pi H)|_{T=M_i}$. If $K_{ia}\gg1$
for any of
the $N_i$, the flavour $\ell_a$ suffers exponentially large washout, whereas
for $K_{ia}\lsim 1$, a substantial fraction of the asymmetry should survive.
In Table~\ref{table:washoutparameters} we list the explicit values for these
washout parameters. From this, we see that indeed, sizeable asymmetries in the
flavour $e$ can persist down to the sphaleron freeze-out. Moreover, we see that
quite the opposite is the case for the flavours $\mu$ and $\tau$, what justifies
our assumptions for calculating the asymmetry that are described
in Section~\ref{sec:calculation} [in particular assumption~(\ref{cond:mutauwashout})].

\begin{table}
\begin{center}
\begin{tabular}{|c|ccc|}
\hline
\diagbox{i}{a} & $e$ & $\mu$ & $\tau$\\
\hline
$1$ & $0.25$ & $2.0\times 10^2$ & $1.8 \times 10^3$\\
$2$ & $0.46$ & $2.3\times 10^2$ & $1.9\times 10^3$\\
$3$ & $0.35$ & $1.7\times 10^2$ & $1.1\times 10^3$\\
\hline
\end{tabular}
\end{center}
\caption{
\label{table:washoutparameters}
The washout parameters $K_{ia}$ for the model specified through the
reference point~(\ref{example:parameters}).
}
\end{table}

The moderately large imaginary parts
of the mixing angles imply some tuning that may be estimated as
$\cosh(\omega_{23})^2\cosh(\omega_{13})^2\cosh(\omega_{12})^2\approx 60$,
what implies that for our example,
the $Y_{ia}$ are generically about a factor eight larger than expected from the standard seesaw relation without a particular alignment.
This amount of tuning can also be explicitly verified by observing cancellations
of individual terms that add up to the largest of the light neutrino masses.
Therefore, finding
viable mixing angles for small $M_i$ requires parametric tweaks.
However, from \fig~\ref{fig:BAU} it
is clear that the situation becomes largely relaxed when allowing for larger
$M_i$.

\begin{figure}[t!]
\begin{center}
\epsfig{file=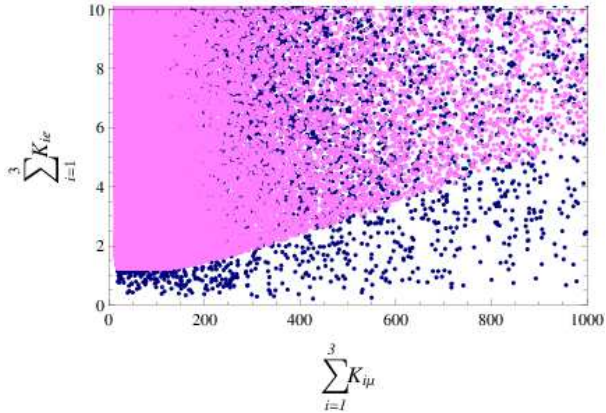,width=8cm}
\end{center}
\vskip-.4cm
\caption{
\label{fig:Kscatter}
Washout parameters $\sum_{i}K_{ie,\mu}$ for a see-saw model with three sterile neutrinos in the mass ratio
$1:2:3$ for $M_1:M_2:M_3$ and the light neutrino mass $m_1=2.5\,{\rm meV}$, over
a flat random distribution for the angles and phases $\delta$, $\alpha_1$, $\alpha_2$, $\omega_{23}$, $\omega_{13}$
and $\omega_{12}$ (dark blue). The same for a model with two sterile neutrinos (light magenta,
implemented here
by setting $m_1=0$, $\omega_{23}=0$ and $\omega_{13}=\pi/2$).
}
\end{figure}

Note that for our example point, we choose a scenario with three sterile neutrinos, whereas it is often
useful to consider the limiting cases where one of these decouples or is simply
not present. For such scenarios, it follows that one of the light neutrinos is massless,
{\it i.e.} $m_1=0$ for normal neutrino mass hierarchy. However, the present scenario of
Leptogenesis is most efficient, if the coupling of the flavour $e$ to the sterile neutrinos is small, to suppress washout, while the couplings $Y_{i\mu}$ and $Y_{i\tau}$ should be large, what
enhances the asymmetry~(\ref{flavoured:asymmetries}). Of course, larger couplings $Y_{ie}$
enhance the asymmetry as well, but this effect is substantially antagonised by the washout effect.
The parameter scan presented in \fig~\ref{fig:Kscatter} illustrates that favourable
parametric
configurations of the Yukawa couplings $Y$ can better be achieved in a scenario with three sterile neutrinos when compared to the
decoupling scenario, where an increase in the couplings $Y_{i\mu}$ automatically implies
a larger lower bound on the couplings $Y_{i e}$.

\paragraph*{Evolution of flavoured asymmetries prior to freeze out.--}

\begin{figure}[t!]
\begin{center}
\epsfig{file=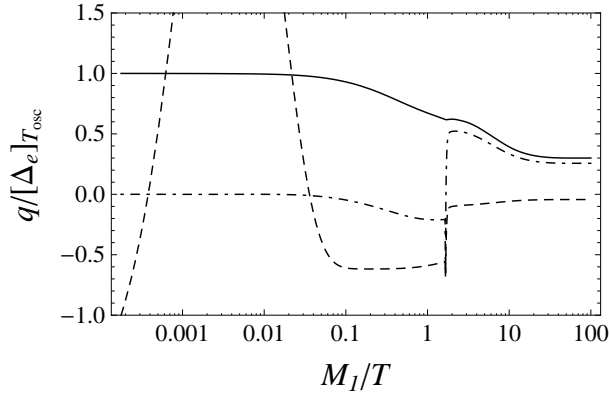,width=8cm}
\end{center}
\vskip-.4cm
\caption{
\label{fig:SterileWO}
Evolution of various combinations of flavoured asymmetries for the reference parameters
and with $M_1=2\times10^4\,{\rm GeV}$ after the initial oscillations
at $z\approx z_{\rm osc}$: $\Delta_e$ (solid), $\Delta_\mu+\Delta_\tau$ (dashed),
$\Delta_{\rm gen}$ (dot-dashed).
}
\end{figure}

In the analysis presented to this end, we follow the evolution of the asymmetry within
the flavour $e$ but not within $\mu,\tau$
because it should be subdominant at the time of freeze out.
The reason is that for the parametric range that we investigate, the flavours $\mu,\tau$
experience a much stronger washout than $e$.
Besides, we neglect helicity asymmetries, that are present within the sterile neutrinos.
At early times, when the Majorana masses are small compared to the temperature, lepton number $L$
is approximately conserved when attributing the charge $L=1$ to a positive-helicity sterile neutrino
and $L=-1$ to a sterile neutrino with negative helicity. The approximate $L$-conservation
during this regime of flavoured washout can be explicitly verified up to the point when the temperature becomes comparable to the mass of the
sterile neutrinos and the LNV washout sets in.

We therefore define the helicity-asymmetry density in the sterile neutrinos as
\begin{align}
\label{helicity:asymmetry}
q_{Ni}=n_{Ni+}-n_{Ni-}\,,
\end{align}
where $n_{Ni\pm}$ are the number densities with helicity $\pm$, and we also introduce
the generalised baryon-minus-lepton-number density
\begin{align}
\Delta_{\rm gen}=\sum\limits_{a=e,\mu,\tau}\Delta_a+\sum\limits_i q_{Ni}\,.
\end{align}
Note that $\Delta_{\rm gen}$ is conserved in the limit $M_i\to 0$, what implies that
it should remain  close to zero for symmetric initial conditions before the temperature
drops below the masses of the sterile neutrinos.

The evolution equations for the flavoured asymmetries are presented in 
Appendix~\ref{appendix:helicity:asymmetries}. In \fig~\ref{fig:SterileWO}, we
present the evolution of particular asymmetries for the reference parameters, where
we choose $M_1=10^5\,{\rm GeV}$. We can indeed verify that the asymmetry $\Delta_e$
is the only one that remains sizeable when $M_1/T\gg1$. In turn the fact that
the generalised baryon-minus-lepton asymmetry $\Delta_{\rm gen}\approx 0$ when
$M_1/T\ll 1$ confirms the approximate conservation of the generalised lepton
charge at early times. Note also that since $Y_{i\mu}$ and $Y_{i\tau}$ are much larger
than $Y_{ie}$, the flavoured asymmetries asymmetries initially satisfy
$|\Delta_{\mu,\tau}|_{T=T_{\rm osc}}\gg|\Delta_e|_{T=T_{\rm osc}}$, while
lepton number conservation implies that
$\Delta_{\mu}+\Delta_{\tau}|_{T=T_{\rm osc}}=-\Delta_e|_{T=T_{\rm osc}}$. The initial
redistribution of the large asymmetries in the $\mu,\tau$ sector to the sterile neutrinos
therefore explains the strong initial incline of $\Delta_\mu+\Delta_\tau$.
Note that it may also be instructive to compare \fig~\ref{fig:SterileWO}
with the schematic diagram explaining the various stages of ARS Leptogenesis
presented in Ref.~\cite{Shuve:2014zua}.

Note that \fig~\ref{fig:SterileWO} also illustrates the main difference
between the original ARS Leptogenesis scenario with sterile neutrinos below the electroweak
scale to the present one with heavier neutrinos. While in the original setup,
total lepton number is conserved above the electroweak phase transition,
implying that $\Delta_{\rm gen}=0$ all the way down to $T_{\rm EW}$, we
see that for the present scenario
the LNV washout processes drive $\Delta_{\rm gen}$ to non-zero values around
$T\approx M_i$\footnote{LNV processes are only important around these times, because
for higher temperatures, sterile neutrinos are produced and destroyed mainly
in lepton-number
conserving scattering-processes, while at lower temperatures, the LNV decays
and inverse decays of sterile neutrinos become Maxwell suppressed.}. As
a result, there is a sizeable residual asymmetry
$\sum_a \Delta_a=\Delta_{\rm gen}\not=0$
within the active leptons
at $T=T_{\rm EW}$, at which point no sterile neutrinos are present any more (such that
$q_{Ni}\equiv 0$). This is
in contrast to the original ARS scenario, where the active asymmetry is balanced by the
helicity asymmetries in the sterile neutrinos,
$\sum_a \Delta_a=-\sum_i q_{Ni}$,
such that $\Delta_{\rm gen}=0$.

In \fig~\ref{fig:fraction},
we also compare the results of the calculation of the freeze-out asymmetries based
on the approximations of Section~\ref{sec:calculation}, where the initial evolution
of the asymmetries in $q_{Ni}$ and $\Delta_{\mu,\tau}$ is neglected, with the results from
the equations of Appendix~\ref{appendix:helicity:asymmetries}. We observe that
in the regimes of either full equilibration of ${\rm R}_e$ (small $M_1$) or no equilibration
(large $M_1$), the calculation based on Section~\ref{sec:calculation} underestimates
the washout at the $5-10\%$ level. This can be attributed to the fact that opposite
asymmetries stored in $\ell_{\mu,\tau}$ get transferred via helicity asymmetries
in the $N_i$ and then partly cancel the asymmetry in $\ell_e$. The helicity asymmetries
are accounted for in Appendix~\ref{appendix:helicity:asymmetries} but not in
Section~\ref{sec:calculation}. Besides, the dip due to incomplete equilibration
of ${\rm R}_e$ extends over a wider range when obtaining the results from
Appendix~\ref{appendix:helicity:asymmetries} because the dynamics of the three different
$N_i$ are resolved individually.

\section{Conclusions}
\label{sec:conclusions}

We have studied the ARS mechanism for sterile neutrinos
above the electroweak scale, what opens up a
substantial region of parameter space for Leptogenesis
based on the type-I seesaw model. While the generation of the
initial flavoured lepton asymmetries proceeds in the same way as
for the original ARS scenario with sterile neutrinos below the electroweak scale,
the calculation of the washout must include LNV processes here, which is
the main difference to the original proposal.

It is however possible to identify regions of parameter space where, while imposing
the observed parameters from active neutrino mixing and oscillations, the initial asymmetry
is large enough and yet the LNV washout small enough yield asymmetries that are
of order of the observed BAU.
In particular, we find that non-degenerate sterile neutrinos with masses
substantially below
$10^9\,{\rm GeV}$ may account for the BAU,
what has not been found to be viable before and allows for new
cosmologically consistent BSM scenarios.
One should note however that the reheat temperature given
by $T_{\rm osc}$ is by orders of magnitude above the
mass scale of the sterile neutrinos [{\it cf.} Eq.~(\ref{zosc})],
unless assuming a mass degeneracy.
Going to lower masses
and reheat temperatures appears to require parametric tuning, either of the Yukawa
couplings or the sterile neutrino masses, that should then be taken to be close to
degenerate. (In turn, the source term that we discuss here generically adds asymmetries to scenarios of resonant Leptogenesis at low temperature
scales.) Another option may be to extend the particle content of the model
in such a way
that the production of sterile neutrinos at the temperature $T_{\rm osc}$ becomes more
efficient. It is interesting to note that the
loop correction to the Higgs mass-square at the parameter point with the lowest
viable sterile neutrino masses considered here ($M_1=4.8\times{10^3}\,{\rm GeV}$) yields
$\sum_i[YY^\dagger]_{ii}M_i^2/(16 \pi^2)\approx (4\times 10^{-2}\,{\rm GeV})^2$, such that the electroweak
scale is not necessarily destabilised by the sterile neutrinos, a desirable feature
also emphasised 
in Ref.~\cite{Khoze:2013oga}.
Future work on this scenario should encompass a more systematic analysis
of the viable parameter space involving three sterile neutrinos
as well as a more accurate determination
of the relevant $CP$-violating rates.

\subsection*{Acknowledgements}
This work is supported by the Gottfried Wilhelm Leibniz programme
of the Deutsche Forschungsgemeinschaft (DFG) and by the DFG cluster of
excellence Origin and Structure of the Universe.

\begin{appendix}

\renewcommand{\theequation}{\Alph{section}\arabic{equation}}
\setcounter{equation}{0}

\section{Including Helicity Asymmetries of the Sterile Neutrinos}
\label{appendix:helicity:asymmetries}

In this Appendix, we extend the equations describing the
washout of the $e$-flavour~(\ref{washout:equil:eqs}) to include as well $\mu,\tau$
and the sterile neutrinos $N_i$. The resulting more detailed description of the
washout is given by
\begin{subequations}
\label{BEqs:withhelicityasymmetry}
\begin{align}
\frac{dq_{\Delta a}}{d z}=&2\sum\limits_i
Y_{ia}Y^\dagger_{ai}\bar W^{\rm NR}_i\vartheta(M_i-T)\left( q_{\ell a}+\frac12 q_H\right)
\\
+&2 \sum\limits_i Y_{ia}Y^\dagger_{ai} \max(\bar W^{\rm NR}_i,\bar W^{\rm rel})
\vartheta(T-M_i)
\left( q_{\ell a}-q_{Ni}+\frac12 q_H\right)\,,
\notag
\\
\label{BEq:Nirelax}
\frac{dq_{Ni}}{d z}=&
2\bar W^{\rm rel}\sum\limits_a Y_{ia}Y_{ai}^\dagger
\vartheta(T-M_i)
\left(q_{\ell a}-q_{Ni}+\frac12 q_H\right)
-2 \sum\limits_a Y_{ia}Y_{ai}^\dagger \bar \Gamma_i^{\rm NR} \vartheta(M_i-T) q_{Ni}
\,,\\
\frac{d q_{{\rm R}e}}{dz}=&
-2\gamma^{\rm fl}\frac{a_{\rm R}}{T_{\rm ref}}
h_e^2\left(q_{{\rm R} e}-q_{\ell e}+\frac12 q_H\right)
\,,
\end{align}
\end{subequations}
where
$\bar\Gamma^{\rm NR}_i=a_{\rm R} z M_i/(16\pi T_{\rm ref}^2)$ is the decay rate of the
sterile neutrinos [note the explicit factor of $2$ in
Eq.~(\ref{BEq:Nirelax}) that accounts for the
${\rm SU}(2)$-multiplicity]. The helicity asymmetry in the sterile neutrinos
$q_{Ni}$ is defined in Eq.~(\ref{helicity:asymmetry}).

The relations between the asymmetries that appear on the left-hand sides of
Eqs.~(\ref{BEqs:withhelicityasymmetry}) to those on the right-hand sides are
derived following
the usual treatment of spectator fields~\cite{Barbieri:1999ma} and are given by
\begin{subequations}
\begin{align}
q_{\ell a}=&\frac{1}{2886}
\left(
\begin{array}{cccc}
-1221 & 156 & 156 & -1245\\
111 & -910 & 52 & 177\\
111 & 52 & -910 & 177
\end{array}
\right)
\left(
\begin{array}{c}
\Delta_e\\
\Delta_\mu\\
\Delta_\tau\\
q_{{\rm R}e}
\end{array}
\right)\,,
\\
q_H=&\frac{1}{481}\left(-37\Delta_e-52\Delta_\mu-52\Delta\tau+45q_{{\rm R}e}\right)\,.
\end{align}
\end{subequations}

In Eqs.~(\ref{BEqs:withhelicityasymmetry}), flavoured and LNV washout rates
are patched together at the point $M_i=T$. We can justify this by the fact that
for smaller temperatures, the flavoured washout rates that are based on
the production rates of relativistic neutrinos derived in Refs.~\cite{Besak:2012qm,Garbrecht:2013bia} should not be applicable, but are
subdominant compared to the LNV rates, that rise steeply when $T$ falls below $M_i$,
{\it cf.} also the Figures that illustrate the sterile neutrino production rate
as a function of temperature in Refs.~\cite{Besak:2012qm,Garbrecht:2013bia}.
The steep incline of the relaxation rate of the sterile neutrinos once these
become non-relativistic is also clearly reflected within \fig~\ref{fig:SterileWO}.

To increase the accuracy of the calculation, it would be desirable to
use results for the production and relaxation of sterile neutrinos that
are also valid between the relativistic and non-relativistic limits.
Methods for calculating the production rate in this most general regime, where
soft and collinear divergences lead to a substantial comlication of the situation, have been
developed in Refs.~\cite{Garbrecht:2013gd,Laine:2013lka}. However, the numerical
procedures are yet numerically challenging, such that a routine evaluation that
could be used for the present purposes is not available yet. Moreover, neither of the
Refs.~\cite{Garbrecht:2013gd,Laine:2013lka} distinguishes between LNV and lepton
number conserving processes to this end. Including this is an important goal for
extending these works in the future.

We also emphasise in this context, that the results for the sterile neutrino production
are for simplicity typically reported for the total number density to 
date~\cite{Besak:2012qm,Garbrecht:2013bia,Garbrecht:2013gd,Laine:2013lka}. Resolving
the single momentum modes in the Boltzmann equations ({\it cf.} Ref.~\cite{Asaka:2011wq})
should improve the accuracy of calculations
in ARS scenarios by order one, but these also increase the numerical efforts substantially.

\end{appendix}

\end{document}